\documentclass[authorversion]{sigchi-ext}
\usepackage[T1]{fontenc}
\usepackage{textcomp}
\usepackage[scaled=.92]{helvet} 
\usepackage{graphicx} 
\usepackage{balance}  
\usepackage{booktabs} 
\usepackage{ccicons}  
\usepackage{ragged2e} 
\usepackage{textpos}
\usepackage{pdfpages}

\def\plaintitle{A Design Philosophy for Agents in the Smart Home} \def\plainauthor{William Seymour}

\def\plainkeywords{Smart Homes, Speculative Design, Design Fiction}

\title{\plaintitle}

\numberofauthors{1}

\author{%
  \alignauthor{%
    \textbf{William Seymour}\\
    \affaddr{University of Oxford} \\
    \affaddr{Oxford, OX1 3QD, UK} \\
    \email{william.seymour@cs.ox.ac.uk} }}

\definecolor{linkColor}{RGB}{6,125,233}
\hypersetup{%
  pdftitle={\plaintitle},
  pdfauthor={\plainauthor},
  pdfkeywords={\plainkeywords},
  bookmarksnumbered,
  pdfstartview={FitH},
  colorlinks,
  citecolor=black,
  filecolor=black,
  linkcolor=black,
  urlcolor=linkColor,
  breaklinks=true,
}

\begin{document}


\copyrightinfo{\scriptsize Permission to make digital or hard copies of part or all of this work for personal or classroom use is granted without fee provided that copies are not made or distributed for profit or commercial advantage and that copies bear this notice and the full citation on the first page. Copyrights for third-party components of this work must be honored. For all other uses, contact the owner/author(s). \\
{\emph{CHI '20 Extended Abstracts, April 25--30, 2020, Honolulu, HI, USA.}} \\
Copyright is held by the author/owner(s). \\
ACM ISBN 978-1-4503-6819-3/20/04. \\
http://dx.doi.org/10.1145/3334480.3375032}


\maketitle

\RaggedRight{} 

\begin{abstract}
  The home is often the most private space in people's lives, and not one in which they expect to be surveilled. However, today's market for smart home devices has quickly evolved to include products that monitor, automate, and present themselves as human. After documenting some of the more unusual emergent problems with contemporary devices, this body of work seeks to develop a design philosophy for intelligent agents in the smart home that can act as an alternative to the ways that these devices are currently built. This is then applied to the design of privacy empowering technologies, representing the first steps from the devices of the present towards a more respectful future.
\end{abstract}




\section{Introduction}
\textit{My home is my castle.} The home is an extremely important place in people's lives, and one of the few places over which an individual truly has control. The home also represents a place where one can withdraw without fear of being watched or judged.

At the same time, we have seen the encroachment of connected intelligent agents into the home. These devices have powerful sensing and computational capabilities, and as a result are able to offer increased safety, automation, and convenience. Devices such as voice assistants are taking on roles previously only described in science fiction, as teachers helping with homework or as carers for the memory impaired. But there is increasing concern that these devices do not primarily act in the interests of their owners, but rather those of developers and manufacturers whose business models revolve around the shaping of behaviour and collection of personal data.

\marginpar{
    \vspace{-60pt}
    \fbox{
        \begin{minipage}{0.925\marginparwidth}
            {\centering
            \textbf{Author Bio}
            \vspace{1pc} \\
            \includegraphics[width=1\columnwidth]{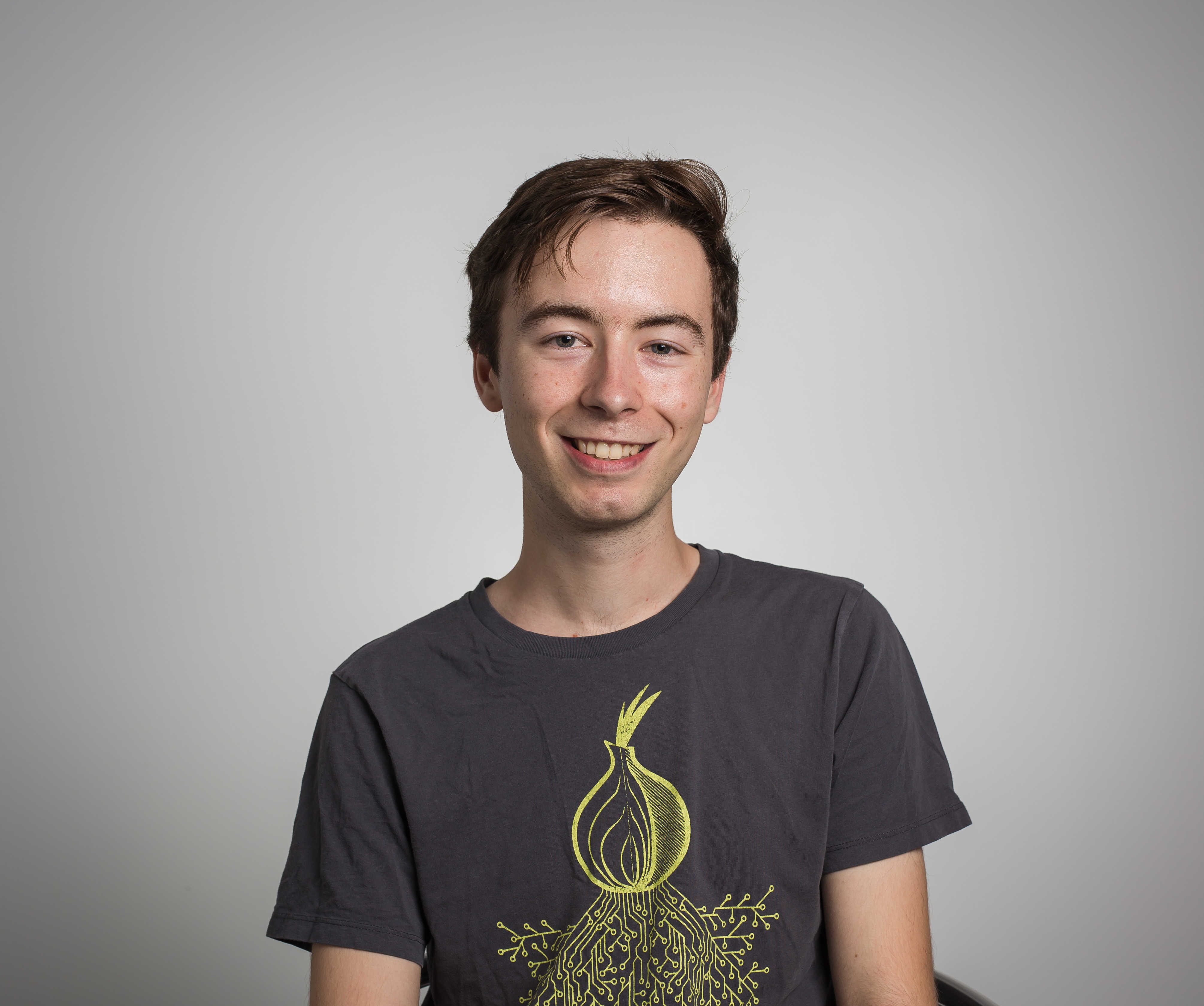}} \\
            \vspace{1pc}
      
            William Seymour is a fourth year DPhil student at the University of Oxford's Department of Computer Science and Centre for Doctoral Training in Cyber Security.
            
            \vspace{1pc}
            
            As part of the Human Centred Computing group he uses speculative design and technology probes to explore alternate design ideologies for intelligent agents in the smart home.
    
        \end{minipage}
    }
}

However this is not the only way that such agents could be designed. To this end, my research explores other ways of designing intelligent agents for the connected smart home. Beginning by mapping out the ethical concerns generated by contemporary smart devices, my work develops an alternate design philosophy of what these devices should do, who they should serve, and how they might better meet users' needs. This is achieved in three distinct but complementary ways:

\begin{enumerate}
    \item \textit{Theoretical}---using philosophical accounts of respect to develop a design philosophy, suggesting ways that intelligent agents might better integrate with people's lives and living spaces
    \item \textit{Speculative}---using speculative design and design fiction to envision scenarios where devices and agents adhere to this design philosophy, showing how the theoretical components inform their design, highlighting how they differ from contemporary products, and critically reflecting on how such products are built
    \item \textit{Empirical}---taking the first steps towards the futures depicted in the speculative elements, my design philosophy will be applied to privacy-empowering technologies in the home, creating prototypes that explore the design space for near-future devices
\end{enumerate}

\section{Background \& Motivation}
\subsection{Smartness as a Marketing, Historical, and Ideological Concept}
As with any loosely defined technological concept (such as cloud computing), perceptions of smartness and what it means for devices to be smart are influenced and guided by a number of interwoven (and often competing) narratives. Marketing narratives of smart devices often present polar views of users and the threats or hazards that they mitigate. Tropes around smartphone usage, for example, are often clustered around the benefits of integration and the perils of dis-integration, pitching the ideal of the connected and productive phone user against fears of becoming addicted or out of touch~\cite{harmon2013stories}. Manufacturers also frequently fall into the trap of `solutionism'~\cite{morozov2013save}, situating devices as solutions to problems without acknowledging the risks and drawbacks that come with them, a process designed to ``bring [detractors] into the fold while keeping [vendors'] central mission of capital accumulation and technocratic governance intact''~\cite{kitchin2015making}.

To an extent, these marketing narratives represent the hopes and dreams of contemporary society, often echoing the depictions of intelligent agents and artificial intelligence in science fiction. Portrayals of future cultures where artificial intelligence enables prosperity and flourishing (e.g. the works of Asimov and others from the `golden age' of science fiction), can be contrasted with more dystopian visions (e.g. works by Gibson, who pioneered the cyberpunk genre of science fiction) where technology is associated with ``human pollution, global destruction, individual control, mental regression, and dehumanisation''~\cite{idier2000science}. This extends to depictions of individual agents or intelligences, such as J.A.R.V.I.S (\textit{Iron Man}, 2008) and HAL 9000 (\textit{2001: A Space Odyssey}, Arthur C. Clark) respectively. Unfortunately, these (literal) narratives, as well as those told by designers and marketers, often come at the expense of more nuanced understandings of how people might or do use technology, pitching practices and values against each other (like the smartphone tropes of integration and dis-integration) and polarising perceptions of those technologies~\cite{harmon2013stories}. 

\subsection{Ethical Concerns in the Smart Home}
Many prior works have investigated the ethical concerns arising from the use of smart home devices, from smart toys~\cite{mcreynolds2017toys}, smart TVs~\cite{malkin2018can}, and smart meter agents~\cite{costanza2014doing,rodden2013home}, to robotic home assistants~\cite{urquhart2018responsible}. Common themes include how automation intended as a convenience can still diminish one's sense of autonomy, and the adverse effects on the relationships and social orders in the home of technology that does not recognise them. Other work has also begun to unpick the subtle long term effects of living with continuously sensing technology designed either explicitly or implicitly as surveillance devices ~\cite{choe2012investigating,oulasvirta2012long}. Research focused more specifically on voice assistants has examined the Media Equation~\cite{reeves1996media} as applied to voice assistants~\cite{mennicken2016s,purington2017alexa}, as well as how families adapt existing structures of authority to voice assistants~\cite{porcheron2018voice}.

A substantial body of work addresses perceptions of and threats to privacy in the smart home. Built on the foundation of Altman's theory of privacy as ``selective control of access to the self''~\cite{altman1975environment}, Palen and Dourish propose that privacy violations take place along boundaries of private/public disclosure, identity, and temporality~\cite{palen2003unpacking}. Nissembaum's theory of contextual integrity similarly demarcate acceptable data flows along the lines of contextual information sharing norms in relation to the subject, sender, recipient, type, and transmission of data~\cite{nissenbaum2009privacy}. But in order to evaluate data flows, users need to be aware of them. Other research has sought to explore potential threats that users are \textit{unaware} of (e.g.~\cite{shklovski2014leakiness}) or lack the required knowledge to fully understand~\cite{tabassum2019don,van2017better}. Sensemaking and similar approaches have shown promise in closing this gap, helping to give users a greater situational awareness of what their devices are recording and sharing about them~\cite{van2018x,van2018need}.

\section{Research Approaches \& Methods}
\subsection{Technology Probes}
Technology probes are fully- or semi-functional prototypes designed be deployed and used over an extended period of time. The data gathered from the probe and it's users is then used to meet a range of research goals: the social science goal of understanding how new technology is used in context, the engineering goal of testing implementations of that technology, and the design goal of generating creative new ways to use technology to meet people's needs and desires~\cite{hutchinson2003technology}.

\subsection{Speculative Design}
Falling under the umbrella of critical design methods, speculative design prompts reflection on contemporary culture and design processes by taking familiar objects and exaggerating select features until they become strange or uncomfortable. By asking questions about the norms that these design artefacts violate, one is forced to ask what kind of society would create such a product, and the extent to which that society differs from or mirrors our own. In this way, speculative design objects act as catalysts for debate about what people really want, rather than suggesting themselves what is desirable~\cite{dunne2013speculative}.

\subsection{Design Fictions}
Like speculative design, design fictions also offer strange or uncomfortable visions of the future, but do so without creating physical artefacts. This allows for the creation of spaces, organisations, and entire worlds that support and contextualise design artefacts, occupying the space between ``the arrogance of science fact, and the seriously playful imaginary of science fiction''~\cite{bleecker2009design}.

\section{Results to Date}
\subsection{(1) Understanding Current Smart Home Devices}
In order to develop a design philosophy for agents in the smart home, it was first necessary to understand the concerns that users have about devices, and how this is related to what makes them `smart'. 

\begin{marginfigure}[-0pc]
  \begin{minipage}{\marginparwidth}
    \centering
    \includegraphics[width=0.9\marginparwidth]{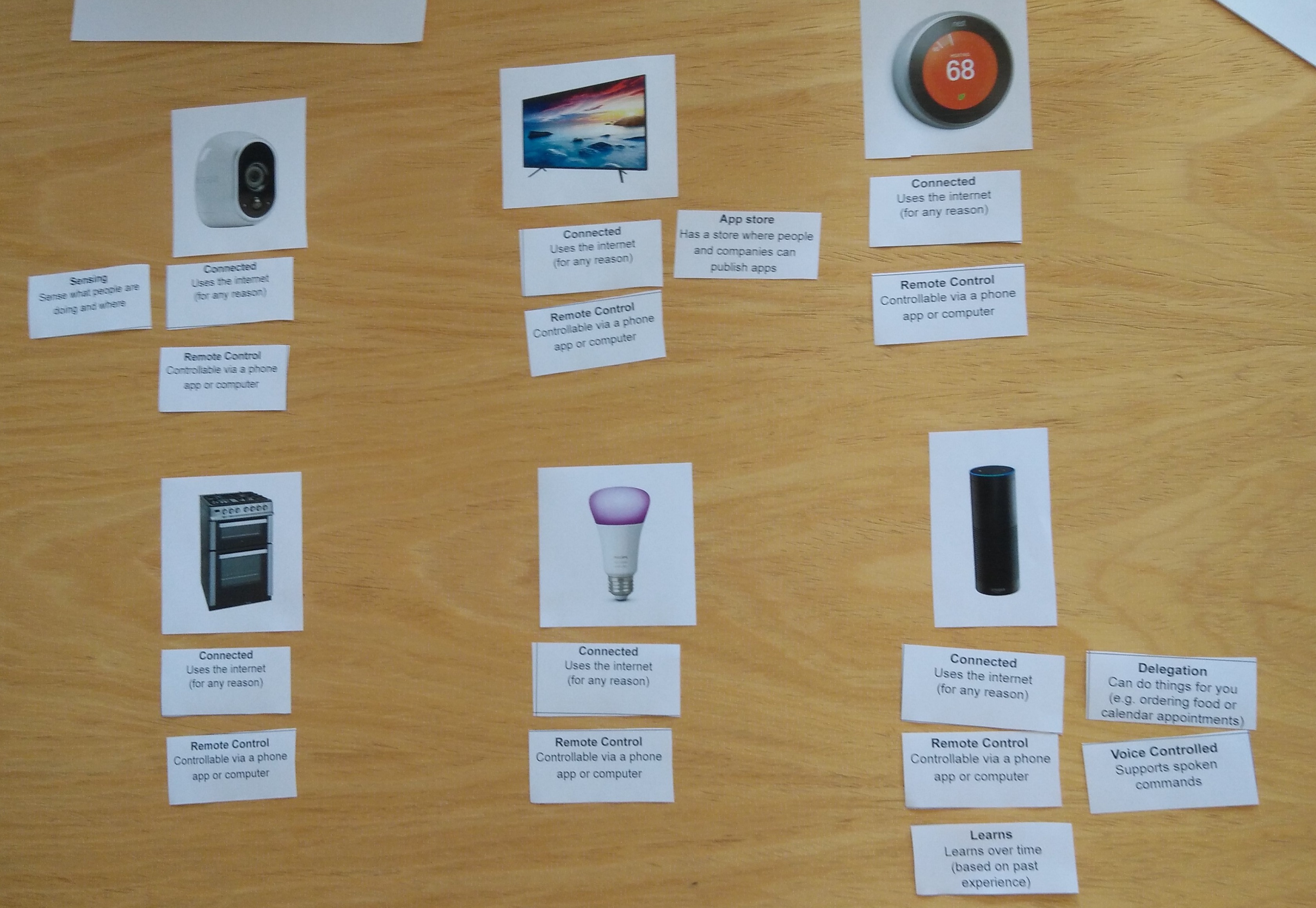}
    \caption{The smart functionality cards as arranged by an interview participant.}~\label{fig:interview}
  \end{minipage}
  \end{marginfigure}

\begin{marginfigure}[4pc]
  \begin{minipage}{\marginparwidth}
    \centering
    \includegraphics[width=0.9\marginparwidth]{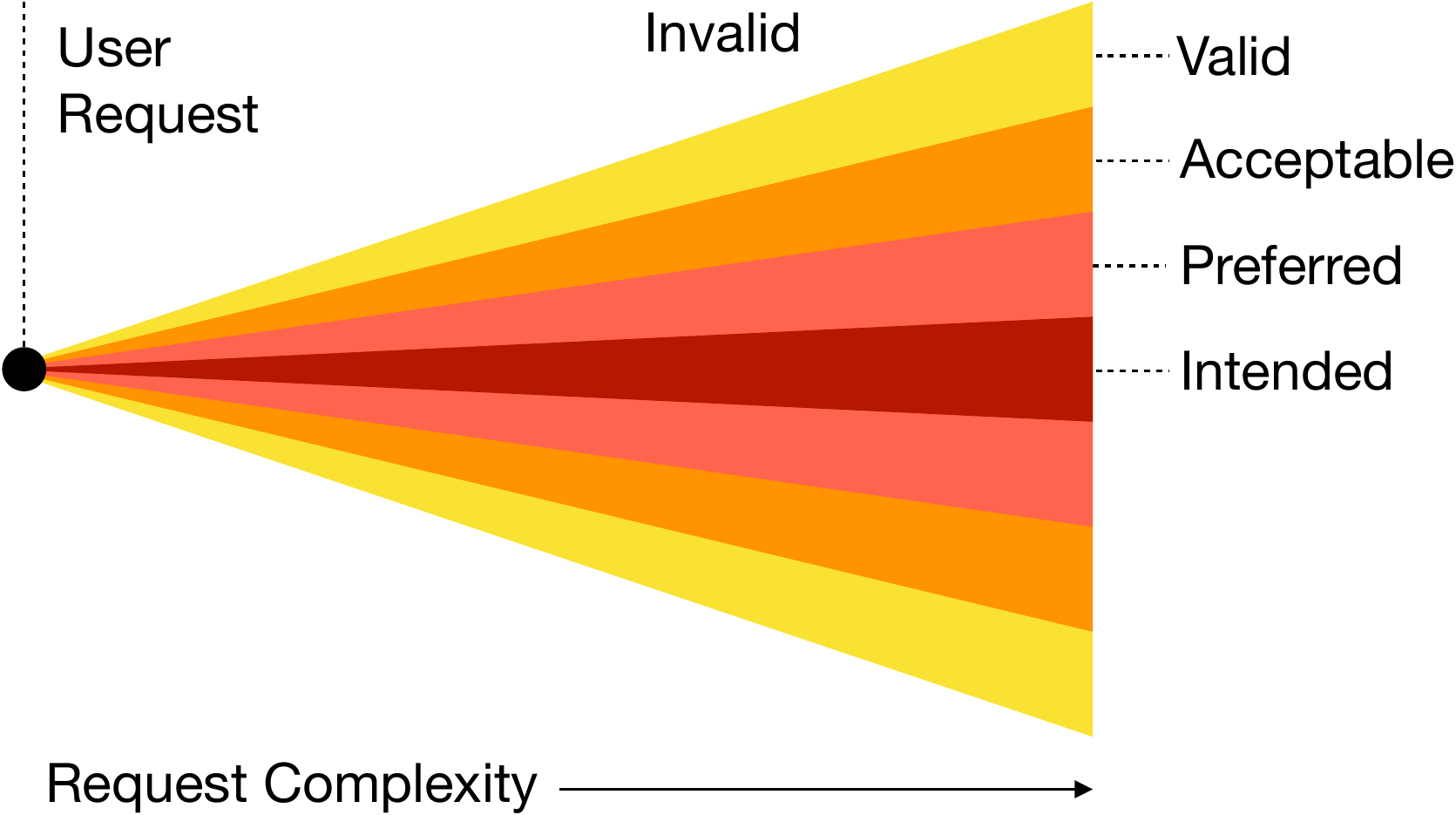}
    \caption{The range of possible responses to under-specified user requests.}~\label{fig:cones}
  \end{minipage}
  \end{marginfigure}

From surveys of smart device owners (n=120), eight different characteristics were identified that were seen as core to devices being `smart'. These then formed the basis for semi-structured interviews (n=15), where participants were introduced to a set of cards with the eight functionalities. After being asked to arrange cards next to devices they thought had that capability (see Figure \ref{fig:interview}), participants were then asked if they would have any concerns about using the device, and whether their concerns would be alleviated by adding/removing functionality cards. In order to go beyond immediate personal reactions to the devices, we then asked them to respond to a set of vignettes depicting smart home contexts (e.g. rental accommodation with devices provided by the landlord).

Thematic analysis of the interviews revealed a variety of interactions between device functionalities and ethical concerns around the use of technology to observe others (particularly children and older adults), transparency around how devices were exercising their capabilities, and the role that devices played as social actors in the home.

This work is currently under review.

\subsection{(2) Using Respect to Inform Device Design}
Having seen how devices, apps, and services balance the needs and interests of users against those of other stakeholders, how might this calculus be changed in order to facilitate the building of trust and comfort? In this work, philosophical accounts or respect are used to precisely characterise and make salient the ways users' needs can be treated by increasingly adaptive and intelligent systems.

Respect is a natural and integral part of human relationships, and an essential skill in social cognition. Respect is also intrinsically related to perceptions of trustworthiness and loyalty. As devices and systems become increasingly advanced, our interactions with them start to resemble human social interactions in several ways. Natural modalities such as speech introduce the possibility for ill-posed or under-specified requests (see figure \ref{fig:cones}), introducing a gulf of interpretation where devices are tasked with inferring what users really need. In this context, we unpack respect into a rich and descriptive typology that describes how future systems might better serve users' needs.

This work is currently under review.

\subsection{(3) Privacy Therapy: What if Your Firewall Could Talk?}
In order to provide an example of how the thesis' alternate design philosophy might be realised, this project presents a speculative prototype of a `respectful Alexa', a smart assistant that is loyal to its users and prioritises their needs over those of it manufacturer and developers. The prototype is engineered using a real Echo unit running a custom Alexa skill. Conversing with the prototype highlights the different ways that such a device might accommodate its user's needs, and is designed to be interoperable with the Aretha technology probe (research item 4 below).

Accompanying the speculative design artefact is an exploration of the technical measures that this class of `respectful' devices could implement. Examples include anonymisation by running extraneous requests (e.g. fetching weather reports from five extra cities), or means by which devices could gain the trust of users (e.g. by stopping heartbeats and analytics data for a short period of time). 

See~\cite{seymour2018loyal,seymour2019privacy} for more detail.

\subsection{(4) Informing the Design of Privacy-Empowering Tools for the Smart Home}
Prior work shows that despite privacy concerns over connected devices in the home, users lack the insight, knowledge, and options essential for taking effective action. To start to understand the potential for new kinds of tools to address these gaps, we developed Aretha, a privacy assistant technology probe that combines a network disaggregator, personal tutor, and firewall to empower users on their home networks.

\begin{marginfigure}[-4pc]
  \begin{minipage}{\marginparwidth}
    \centering
    \includegraphics[width=0.9\marginparwidth]{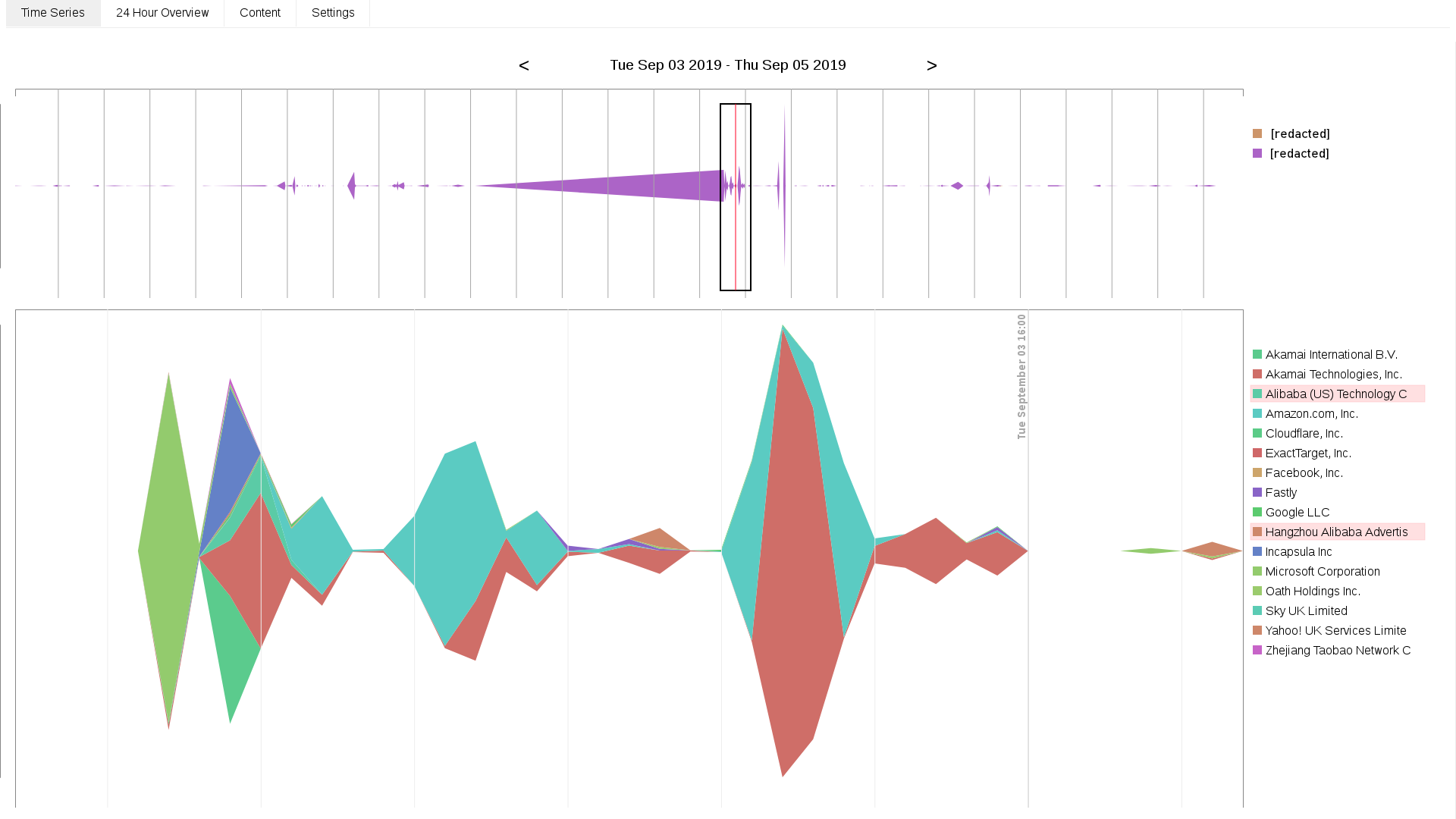}
    \caption{Time series visualisation in the Aretha probe. Shows traffic volume by company and by devices over a 24 hour period.}~\label{fig:aretha}
  \end{minipage}
\end{marginfigure}

To better understand how this combination of resources could allow users to gain awareness of data disclosures by their devices, form educated privacy preferences, and control unwanted data flows, Aretha was deployed in three households over six weeks. During this time participants were instructed to keep track of what was happening on the visualisations (see figure \ref{fig:aretha}) and follow the content in the curriculum. Towards the end of the study the firewall control allowed them to block network traffic between specific devices and companies of their choosing.

The probe encouraged participant households to talk together about security and privacy, and lines of questioning evolved from \textit{where} devices were sending data to \textit{why} those destinations required it (and what were they doing with it). While the firewall ultimately did not meet user's needs, it did prompt them to find their own controls and suggested strategies to improve similar tools in the future.

This work is currently under review.

\section{Next Steps}
\subsection{(5) The Power of Giving Devices a Voice}
Speech activates the same locations of the brain regardless of whether its origin is human or machine. As voice assistants become more sophisticated and more devices support voice control, it is important to understand the effects this might have on those using these technologies. This paper will explore the presence of cognitive shifts that might result from long term use of devices that present themselves as human-like.

A survey of voice assistant users will be used to explore the interplay between three axes along which users perceive their voice assistants as: mechanical devices (inc. reliability, confidence, and consistency); social actors (how interactions relate to social rules and norms governing interpersonal relationships; and anthropomorphism (how much users ascribe human characteristics and intents to voice assistants).
    
From this I hope to discover how these axes interrelate, as well as how they correlate with ownership factors (e.g. length of ownership). It is expected that this will highlight potential concerns around deploying human-like devices, as well as suggesting strategies that can be taken to mitigate the potentially harmful effects of the Media Equation in voice assistants.

\subsection{(6) Voice Assistants with Multiple Personalities}
When you buy a voice assistant, you receive a physical object. The warranty covering the assistant applies to this physical object, but what is really being purchased is access to the agent that `inhabits' the plastic and silicon. The intelligent agents that drive current commercial offerings are designed to speak and act like humans, often engendering feelings of social presence in users~\cite{cho2019hey}. Inspired by Japanese Shinto spiritual practices, this design fiction imagines voice assistants as hosts for different spirits concerned with human beings---kami. These assistants (mostly) want us to be happy; if they are treated properly they can bring us numerous benefits such as automating tasks for us, accessing information, or providing help and advice (known as the harmonious soul), but if they are disregarded then they can sow discord and cause trouble (known as the wild soul).

Through this lens I hope to contrast these fluid and sometimes capricious responses with contemporary notions of voice assistants as polite and subservient, as well as highlighting the gulf of interpretation in a (pseudo) real-world context. The design fiction asks us to imagine what kind of a society might adopt these kami assistants, and how they might adapt to them as they change between harmonious and wild.

\subsection{(7) Informing Data Protection by Design and Default in Future Smart Homes}
From its beginnings as a means of self-regulation, the privacy by design paradigm has recently been enshrined in the GDPR as `data protection by design and default' (Art. 25). Properly enacting the spirit of these regulations in smart home devices requires an understanding of several areas that manufacturers have traditionally paid little attention to, such as the social dynamics surrounding shared or communal devices, and the use of technology by and for vulnerable groups.

To address this, I am part of a project funded by the UK Information Commissioner's Office that will design, prototype, and evaluate a number of novel devices that address the above issues. With the aim of informing the design of the smart home of the future, we will also explore the ways that hardware and software design can be used to raise awareness of the legal rights of data subjects, and how these might be customised to fit users' needs.

See~\cite{kraemer2019informing} for more information.

\section{Expected Contributions}
\begin{itemize}
    \item Exploration of ethical concerns caused by current devices, highlighting the key tensions at play in the context of the connected home. Comprises research items (1) and (5).
    \item An alternate design philosophy for future intelligent agents in the smart home, illustrated by examples of speculative and fictional device designs that adhere to this philosophy. Comprises research items (2), (3), and (6).
    \item Design and implementation of near-future devices that bridge the gap between contemporary products and the design artefacts previously described. This will be done through the creation of prototypes that are then evaluated by users and industry practitioners. Comprises research items (4) and (7).
\end{itemize}

\section{Acknowledgements}
This work is supported by EPSRC through the PETRAS subproject \textit{Respectful Things in Private Spaces} (grant number N02334X/1), and the ICO grants program (2018-2019).

\bibliographystyle{SIGCHI-Reference-Format}
\bibliography{main}






\end{document}